\documentclass[12pt]{iopart}
\usepackage{graphicx}
\usepackage{caption}
\usepackage{subfig}
\usepackage{wrapfig}
\expandafter\let\csname equation*\endcsname\relax
\expandafter\let\csname endequation*\endcsname\relax
\usepackage{amsmath}
\usepackage{amssymb}

\begin{document}

\title[Mode-coupling points to functionally important sites]{Mode-coupling points to functionally important residues in Myosin II}

\author{O. Varol$^{1\dagger,2}$, D. Yuret$^{1\dagger}$, B. Erman$^{1\dagger}$, A. Kabak\c c\i o\u glu$^{1*}$}

\address{$^1$ Colleges of Engineering$^\dagger$ and Sciences$^*$, Ko{\c c} University, Sar{\i}yer, 34450, {\. I}stanbul, Turkey \\
		 $^2$ School of Informatics and Computing, Indiana University, Bloomington IN, USA }
\ead{akabakcioglu@ku.edu.tr}

\begin{abstract}
Relevance of mode coupling to energy/information transfer during
protein function, particularly in the context of allosteric
interactions is widely accepted. However, existing evidence in favor
of this hypothesis comes essentially from model systems. We here
report a novel formal analysis of the near-native dynamics of myosin
II, which allows us to explore the impact of the interaction between
possibly non-Gaussian vibrational modes on fluctutational dynamics. We
show that, an information-theoretic measure based on mode coupling
{\it alone} yields a ranking of residues with a statistically
significant bias favoring the functionally critical locations
identified by experiments on myosin II.

\end{abstract}

%\submitto{\PB}
\maketitle

\section{Introduction}
Fluctuation based analysis of protein dynamics has long proven to be
an invaluable tool for investigating the interplay between protein
dynamics and function~\cite{bahar2005coarse,berendsen2000collective}.

Despite the past success enjoyed by the bead-and-spring-type linear
models (such as, elastic, Gaussian, and anisotropic network models),
it is well known that, both experimental evidence and simulations
reveal strong departure from purely Gaussian (harmonic) behavior at
physiological temperatures~\cite{tilton1992effects}. Deviations from
harmonicity are most pronounced in slow, collective modes which are
significant, for example, in the context of vibrational absorption
spectrometry~\cite{roitberg1995reports}, dimensional
reduction~\cite{Hayward1994Harmonic,kitao1999investigating}, and the
role of hydration effects~\cite{nakagawa2008hydration}.

On the other hand, the decomposition of MD fluctution data into
independent, possibly anharmonic modes is only the first step in an
infinite cascade of corrections that bridge the gap between the
dynamics of actual proteins and Gaussian models. The contribution of
higher-order corrections signifies the degree to which the
experimental/computational free energy landscape fails to conform to a
representation composed of independent modes (harmonic or
anharmonic). In other words, they are ``mode-coupling'' corrections
which yield valuable information on means of energy transfer and
associated correlated activity within the
protein~\cite{Moritsugu2000Vibrational,xie2000long,Leitner2008Energy,Piazza2009Longrange,piazza2010breather,gur2010quasi}.
Characterization of the confirmational population sampled by
near-native dynamics is believed to be the key to understanding the
functioning of allosteric
proteins~\cite{monod1963allosteric,kern2003role}, if not
all~\cite{tsai2008allostery}. The interactions between vibrational
modes play an essential role in shaping this population.

Recently, we introduced a systematic mathematical analysis of the
fluctuational data (for example obtained from full-atomistic
simulations), that naturally distinguishes the anharmonic and
mode-coupling contributions to the free
energy~\cite{Kabakcioglu2010IOPscience}. Here, we combine this
analytical formulation with computer simulations of the near-native
dynamics of myosin II and demonstrate that the mode-coupling {\em
  alone} highlights functionally critical sites of this allosteric
protein. The relevance of coupling between vibrational modes in the
context of allosteric transitions in myosin II was also pointed out in
an earlier work~\cite{zheng2009coupling}.

The paper is organized as follows:
Section~\ref{section:modal_expansion} describes the theoretical
framework used for isolating the contribution of mode-coupling from
other anharmonic effects in the MD fluctuation data;
Section~\ref{section:residue_score} discusses how the formulation
above can be used to select out residues that are highlighted by
mode-coupling; Section~\ref{section:myosin_dynamics} introduces the
motor protein myosin II which we use here as a test case;
Section~\ref{section:MD} gives the details of the molecular dynamics
(MD) simulations performed on myosin II; Section~\ref{section:results}
reports and Section \ref{section:discussion} discusses our results.

\section{Materials and Methods}
\subsection{Modal expansion and beyond}
\label{section:modal_expansion}
Our raw data is the time-series for the space coordinates of the
$\alpha$-carbons obtained from a full-atomistic molecular dynamics
(MD) simulation whose details are given in
Section~\ref{section:MD}. Using the MD trajectory, we derive a
multivariate probability distribution function $p(\Delta \mathbf{R})$,
where $\Delta R_i$, $\Delta R_{i+N}$, and $\Delta R_{i+2N}$ with
$i=1,...,N$ are the deviations from the mean position along the
coordinate axes $x$, $y$, and $z$, respectively, of the $i$-th
$C^\alpha$ atom in a protein with $N$ amino acids. The covariance
matrix $\Gamma = \left \langle \Delta \mathbf{R} \Delta \mathbf{R}^T
\right \rangle$ is then used to transform the coordinate system by
means of a scaling and a rotation into the modal space: $\Delta
\mathbf{r} = \Gamma^{-1/2} \Delta \mathbf{R}$. For a purely harmonic
system, the resulting distribution function is given by
\begin{equation}
\label{eq:gaussian}
f(\Delta \mathbf{r}) = \prod_i \frac{\exp[-\Delta r_i^2/2]}{\sqrt{2\pi}}
\end{equation}
while deviations from Eq.(\ref{eq:gaussian}) due to anharmonicity and
mode-coupling are observed in proteins, as mentioned before.

\subsection{Hermite expansion}
Building on an earlier proposal~\cite{yoon1974moments}, we recently developed an
analytical formalism that naturally extends Eq.(\ref{eq:gaussian})
into the regime where harmonicity breaks
down~\cite{Kabakcioglu2010IOPscience,gur2010quasi}. In this framework,
$f(\mathbf{\Delta r})$ is expressed as an infinite sum:
\begin{eqnarray}
\label{eq:hermite_expansion}
f( \mathbf{\Delta r}) &=& \frac{1}{\sqrt{(2\pi)^{3N}}} e^{-\sum_i \Delta r_i^2 /2} \bigg[ 1 + \sum_i \sum_{\nu =3}^\infty c_\nu^i H_\nu( \Delta r_i)\nonumber \\
&+& \sum_{i\neq j}\sum_{\nu =3}^\infty \sum_{p=1}^{\nu -1} c_{p,\nu-p}^{ij} H_p (\Delta r_i) H_{\nu-p}(\Delta r_j) \nonumber \\
&+& \sum_{i\neq j\neq k}\cdots\bigg]
\end{eqnarray}
where $H_i$ is the Hermite polynomial of rank $i$. The choice of the
Hermite basis ensures that the expansion coefficients are given by
$c_\nu^i = \langle H_\nu ( \Delta r_i ) \rangle/\nu!$ and
$c_{p,\nu-p}^{ij} = \binom{\nu}{p} \langle H_p( \Delta r_i) H_{\nu
  -p}( \Delta r_j ) \rangle/\nu!$, where $\langle\cdot \rangle$
denotes the time average evaluated over the MD
data. Ref.\cite{Kabakcioglu2010IOPscience} describes how the symmetry
properties of Hermite tensor polynomials can be exploited to reduce the
computational complexity associated with estimating these coefficients
from the MD trajectory.

The leading term in Eq.(\ref{eq:hermite_expansion}), which is
identical to Eq.(\ref{eq:gaussian}), corresponds to a purely harmonic
dynamics and is referred as $f_0$ here. This zeroth-order form is the
basis for many protein fluctuation models
\cite{bahar1997direct,atilgan2001anisotropy,yogurtcu2009statistical}.
Remaining terms within the square brackets in
Eq.(\ref{eq:hermite_expansion}) reflect all possible corrections due
to non-Hookian modes, as well as pairwise, threesome, and higher-order
mode-mode interactions. We wish to focus on the impact of
mode-coupling in our study, therefore our first goal is to distinguish
the contributions that yield anharmonic (and still independent) modes
from those that are due to the interactions among such modes. Here, we
will prefer to the former as {\it marginal anharmonicity}, since this
contribution is uniquely determined by the deviations of the marginal
distributions $f(\Delta r_i) = \int \prod_{j\ne i}\,dr_j f(
\mathbf{\Delta r})$ from Gaussian.

To this end, let $f_1$ refer to the best possible description of the
data under the assumption of marginal anharmonicity:
\begin{eqnarray}
\label{eq:hermite_f1}
f_1(\mathbf{\Delta r}) &=& \frac{1}{\sqrt{(2\pi)^{3N}}} e^{-\sum_i \Delta
  r_i^2 /2} \prod_i \bigg[ 1 + \sum_i \sum_{\nu =3}^{\nu_{max}} c_\nu^i
  H_\nu( \Delta r_i) \bigg]
\end{eqnarray}
where $\nu_{max}$ is a cut-off degree imposed by practical
considerations (see Section~\ref{section:myosin_dynamics}). This
approximation to the conformational distribution function obtained from
near-native dynamics yields exact single-mode (marginal)
histograms in the limit
$\nu_{max}\to\infty$. Note that, $f_1$ is fully specified by the
coefficients $\{c_\nu^i\}$. Nevertheless, marginal anharmonicity is
reflected at all orders in $f(\mathbf{\Delta r})$ (i.e.,
$c_{p,\nu-p}^{ij}$ and higher-order coefficients are typically
nonzero). At first sight, these high-order contributions may be
confused with mode-coupling since they are in the form of a product
involving multiple vibrational modes. However, it is transparent from
Eq.(\ref{eq:hermite_f1}) that, the information on mode-mode interactions is
contained in {\it everything but $f_1$}.

\subsection{Mode-coupling based ranking of residues}
\label{section:residue_score}
It is tempting at this point to attempt to identify pairs of modes
which interact strongly and/or have the most impact on protein
function. Numerous studies in this spirit can be found in the
literature (see, e.g.,
~\cite{zheng2009coupling,zheng2005identification}. However,
interpreting such data usually requires an understanding of the
functional dynamics and does not immediately relate to experiments.
Furthermore, a pairwise interaction picture is incomplete in the
current context, because the corrections to the fluctuational free
energy are not additive in mode pairs. In other words, higher-order
contributions exist.

Instead, we here focus directly on the critical residues of the
protein which are highlighted by mode-coupling at all orders. This
kind of information is not only easier to compare with available
experimental data (such as site-directed mutation scans), but, as it
turns out, it is also computationally cheaper to access. As is evident
from Eq.(\ref{eq:hermite_expansion}), estimating mode-coupling
corrections per mode pair involves calculating second and higher-order
coefficients $c_{\nu\eta\dots}^{ij\dots}$ associated with the
individual mode pair, repeated for $N \choose 2 $ pairs; a CPU
demanding task. The cumulative effect of mode coupling, however, is
already available in the difference between $f$ and $f_1$. This
information can be projected onto the protein's sequence axis by the
procedure outlined in Section~\ref{section:res_impact}. The outcome is
a score profile for each amino acid in the protein, reflecting the
degree to which their near-native fluctuations are modulated by mode
coupling.

\subsection{Identifying per residue impact of mode-coupling}
\label{section:res_impact}

In order to identify the residues highlighted by marginal
anharmonicity and mode coupling, separately, we back-project the
distributions $f_0$ and $f_1$ onto the space of $C^\alpha$ atomic
coordinates:
\begin{eqnarray}
\label{eq:ftilde}
p_{0,1}(\mathbf{\Delta R}) &=&  f_{0,1}(\mathbf{\Delta r}(\mathbf{\Delta R}))/\sqrt{\mbox{det}\ \Gamma}\ .\nonumber
\end{eqnarray}
$p_0$ and $p_1$ are approximations, at two different levels (Gaussian
and marginally anharmonic), to the original distribution
$p(\mathbf{\Delta R})$ obtained from the MD trajectory.

Next, we consider the marginal distributions $p(\mathbf{\Delta R_i}) =
\int \prod_{j\ne i} \mathbf{dR_j}\ p(\mathbf{\Delta R})$ for
individual coordinates $C^\alpha_i$ and measure the Kullback-Leibler
(KL) divergence~\cite{kullback1951information}, $d_{KL}$, between $p$
and $p_1$, as well as between $p_1$ and $p_0$ for a given residue
$i$. The former distance yields quantitative information on the extent
to which mode-coupling governs fluctutations of the given coordinate
$C^\alpha_i$, while the latter yields a similar measure as regards to
marginal anharmonicity. For distributions $p$ and $q$ of a continuous
random variable $x$, KL-divergence is defined to be the integral
\begin{equation}
d_{\mathrm{KL}}(p\|q) = \int_{-\infty}^\infty p(x) \ln
\frac{p(x)}{q(x)} \, {\rm d}x\ .
\end{equation}
The integration steps involved in the KL divergence estimation require
the discrete probability distributions obtained from the MD data to be
smoothened out into continuous functions. To this end, we use kernel
density estimation (KDE)
\cite{rosenblatt1956remarks,parzen1962estimation} which yields a
continuous probability distribution $\hat{p}(x)$ from a set of samples
$\{x_i\}$ as
\begin{equation}
\hat{p}_h(x) = \frac{1}{nh} \sum_{i=1}^n{K\left ( \frac{x-x_i}{h}
  \right )}
\end{equation}
where $K$ is the kernel function (chosen to be Gaussian) and $h$ is
the bandwidth parameter (determined by the method in
Ref.~\cite{Bernard1986Density}).

The total impact for a given residue is taken to be the sum of the
KL-divergence values for its three spatial coordinates:
\begin{eqnarray}
S^{mc}_i = \sum_{\alpha=x,y,z}\,d_{KL}[p(\mathbf{\Delta
    R_{i,\alpha}}) \| p_1(\mathbf{\Delta R_{i,\alpha}})] \ \ , \label{eq:res_score_a}\\
S^{ma}_i = \sum_{\alpha=x,y,z}\,d_{KL}[p_1(\mathbf{\Delta
    R_{i,\alpha}}) \| p_0(\mathbf{\Delta R_{i,\alpha}})]\ . \label{eq:res_score_b}
\end{eqnarray}
Above, ``$mc$'' and ``$ma$'' stand for ``mode coupling'' and ``marginal
anharmonicity'', respectively. As a reference, we also consider the
mean residue displacements (akin to experimental B-factors) measured
by the variation
\begin{equation}
\sigma_i^2 \equiv \int \mathbf{\Delta R_i}^2 p(\mathbf{\Delta R_i})
  d\mathbf{\Delta R_i}
\label{eq:fluc_score}
\end{equation}
 as the fluctuation-based score for a residue. Below, we apply this
 analysis to the MD data from myosin II, a molecular motor protein,
 and compare the performance of the above ranking schemes in
 distinguishing functionally significant locations on the protein.
 Note that, once the MD data is available, estimation of the residue
 scores above is a mechanical process, without any tuning paramaters.

\subsection{Myosin II}
\label{section:myosin_dynamics}
\textit{Dictyostelium discoideum} myosin II is an allosteric protein
which has been extensively studied both experimentally and
computationally. It is an actin-binding molecular motor
protein crucial for various biological processes, such as, cell
movement, muscle contraction in higher organisms, membrane transport
and several signaling pathways. Among the 35 known classes of myosin,
13 appear in Human~\cite{Sweeney2010Structural}. The motor domain of
myosin II shown in Fig.\ref{fig:FunctionalSites} goes through
conformational changes at each stage of its four-stroke catalytic
cycle which converts the chemical energy derived from ATP hydrolysis
into mechanical work. The results presented below are obtained from
MD simulations of the structure PDB:1VOM~\cite{Smith1996Xray}, where
an ADP is bound on the protein.

\begin{figure}[h]
  \includegraphics[width=\textwidth]{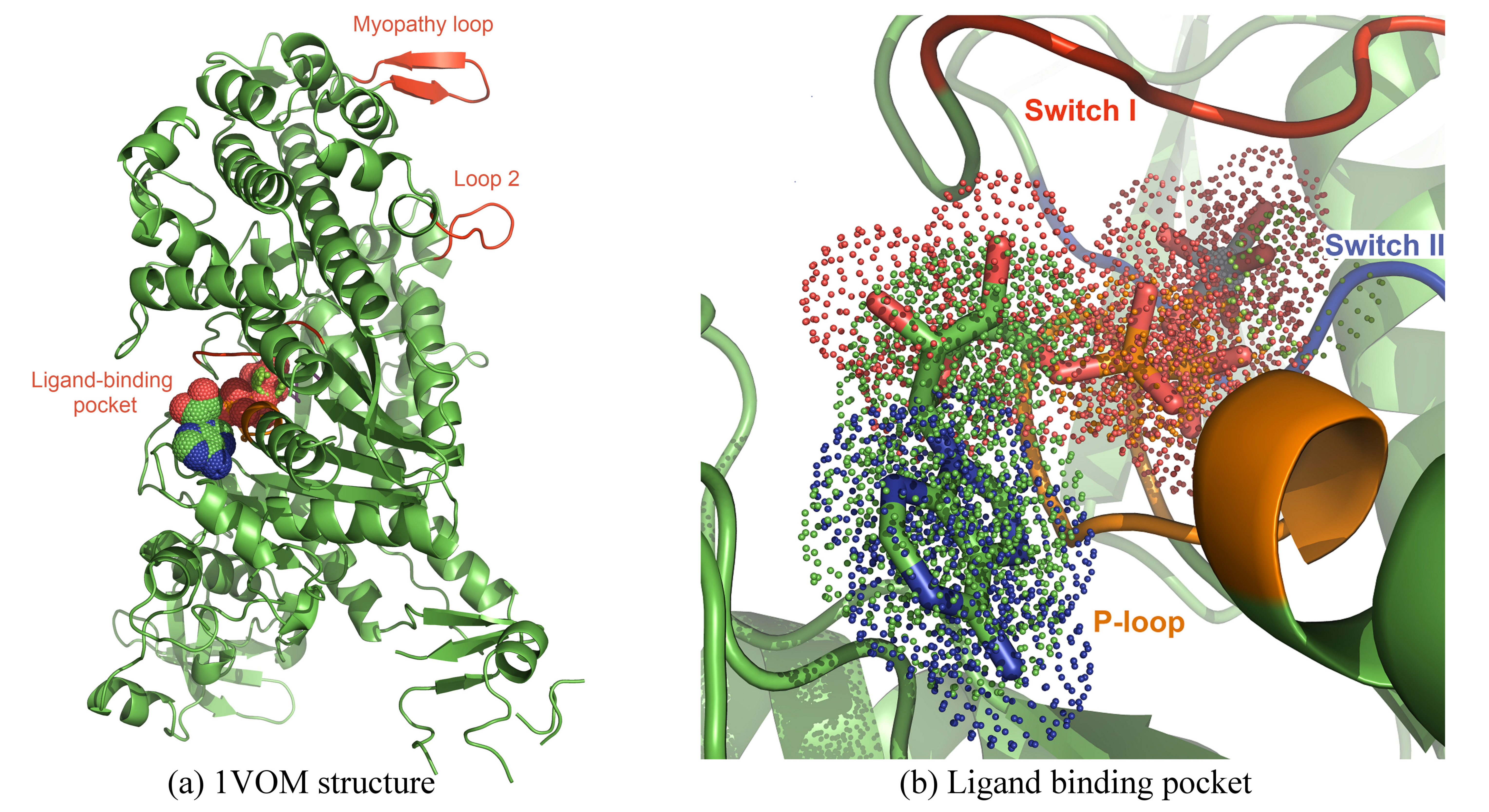}
  \label{fig:FunctionalSites}
  \caption{Functional sites of myosin II (PDB:1VOM). The full
    structure of the motor domain is shown on the left (a).  Actin-
    and ADP-binding loops are indicated in red. On the right, the
    ligand binding pocket is shown in greater detail (b). The ADP
    molecule in the middle is surrounded by functional elements Switch
    I (red), Switch II (blue) and P-loop
    (orange). (Fig. 1a)}
  \label{fig:FunctionalSites}
\end{figure}

There exist several, well-studied functionally relevant locations on
the structure. Switch II and P-loop shown in
Fig.\ref{fig:FunctionalSites} are known to control the MgADP release
mechanism~\cite{Smith1996Xray}. Structural changes during the
characteristic cleft closure motion in the motor domain is believed to
be related to activity in Switch I region, which opens the binding
pocket and modifies the relative placement of the P-loop and Switch II
regions~\cite{reubold2003structural}. The actin-binding pocket of the
structure is composed of the Myopathy loop and Loop-2, shown in
Fig.\ref{fig:FunctionalSites}. The interaction between Loop-2 and the
negatively charged parts of the actin is also
documented~\cite{Spudich1994How}.

Below, we perform the proposed fluctuational analysis on the motor
domain of myosin II and identify the residues whose fluctuations are
most significantly modified by mode-coupling. We find that, there is a
statistically significant correlation between these and the functional
regions mentioned above, as well as a subset of critical residues of
the protein determined by experimental methods (such as point
mutations).

\subsection{MD simulations and the eigenmodes}
\label{section:MD}
The structure is composed of 730 residues and the ligand whose atomic
coordinates (the initial configuration of the MD simulation) were
extracted from the PDB database.  The MD simulations were carried out
using NAMD 2.7 software package \cite{NAMD} with CHARMM27 force field
\cite{CHARMM2009} in explicit solvent (water) at 310 K. Langevin
dynamics was used to control the temperature and the pressure in an
NPT ensemble. A water box with a 15$\AA$ cushion and periodic boundary
conditions were applied. The integration time step in the simulation
was selected as 1 fs for both non-bonded and electrostatic forces and
no rigid bonds were used. The trajectory was captured every 50 fs
within several windows of $\sim 2$ ns duration, for a total run of 10
ns.

Note that, a much longer simulation time would be required to observe
the functional dynamics of the protein. The purpose of our simulation,
however, is merely to monitor the fluctuations and to gather
sufficient data on the non-Gaussian nature of the conformational
distribution. The procedure may be crudely likened to recording a
short bike ride and then analyzing the small displacements of various
elements in order to identify the components that are critical in
transfer mechanical energy (except, thermal fluctuations are
significantly more influential in the current system.)

\begin{figure}[h!]
  \centering
  \includegraphics[width=0.75\textwidth]{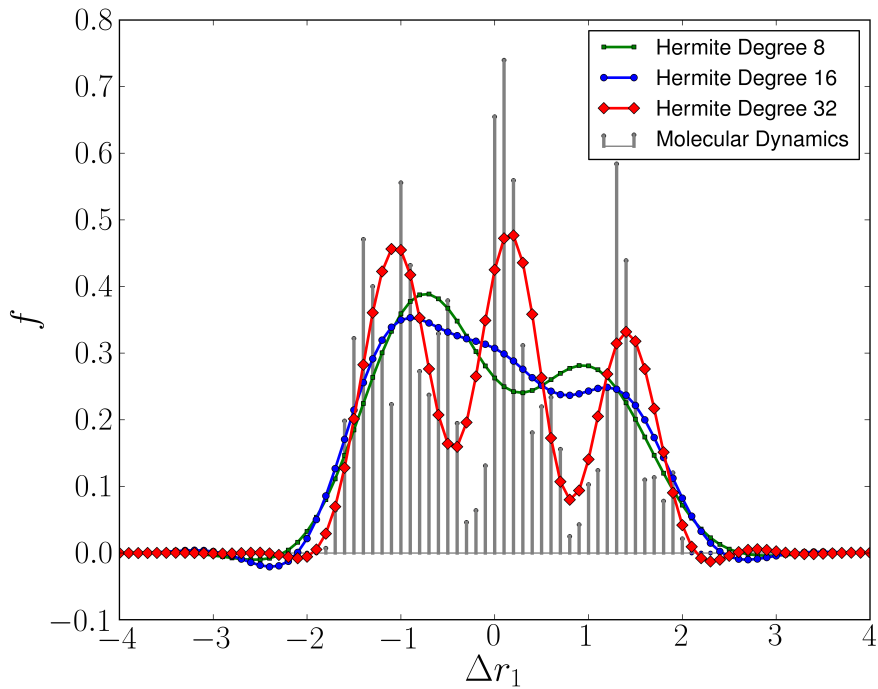}
  \caption{A comparison of the slowest mode's amplitude distribution
    for different choices for the maximum Hermite degree considered in
    Eq.(\ref{eq:hermite_f1}). The cut-off degree $\nu_{max}=32$ was
    determined according to the criterion that the marginal
    distributions for all modes are captured with an accuracy same as
    above or better.}
  \label{fig:hermite_degree}
\end{figure} 

The MD output was analysed in several time windows. Let us consider
the steps of the analysis on the first 2 ns of the simulation data
after equilibration, where snapshots taken 0.5 ps apart amount to
$N=4000$ data points: We first construct the vector $\Delta
\mathbf{R}^{(i)}$ of the $C^\alpha$ positions in each snapshot, with
$i=1,\dots,N$. We next calculate the covariance matrix $\Gamma$ given
in Section~\ref{section:modal_expansion}, and identify the modal
coordinates $\Delta \mathbf{R}^{(i)}$ for the 2184 fluctuation modes
(out of $730\times 3=2190$ degrees of freedom, excluding six
associated with the center of mass translation/rotation). At this
point, the zeroth-order approximant $f_0$ to $f(\Delta
\mathbf{r}^{(i)})$ (and to $f(\Delta \mathbf{R}^{(i)})$, through the
inverse transform) is already available. Next, we find the best
marginally anharmonic description of the data, $f_1$, given by
Eq.(\ref{eq:hermite_f1}). This is done by estimating $c_\nu^i$ in
Eq.(\ref{eq:hermite_f1}) as averages over the MD snapshots, up to a
sufficiently high cut-off degree $\nu_{max}=32$ which is obtained
empirically (see Fig.\ref{fig:hermite_degree}). Finally, we calculate
the residue scores using
Eqs.(\ref{eq:res_score_a}-\ref{eq:fluc_score}).

\begin{figure}[h!]
  \centering
  \includegraphics[width=0.75\textwidth]{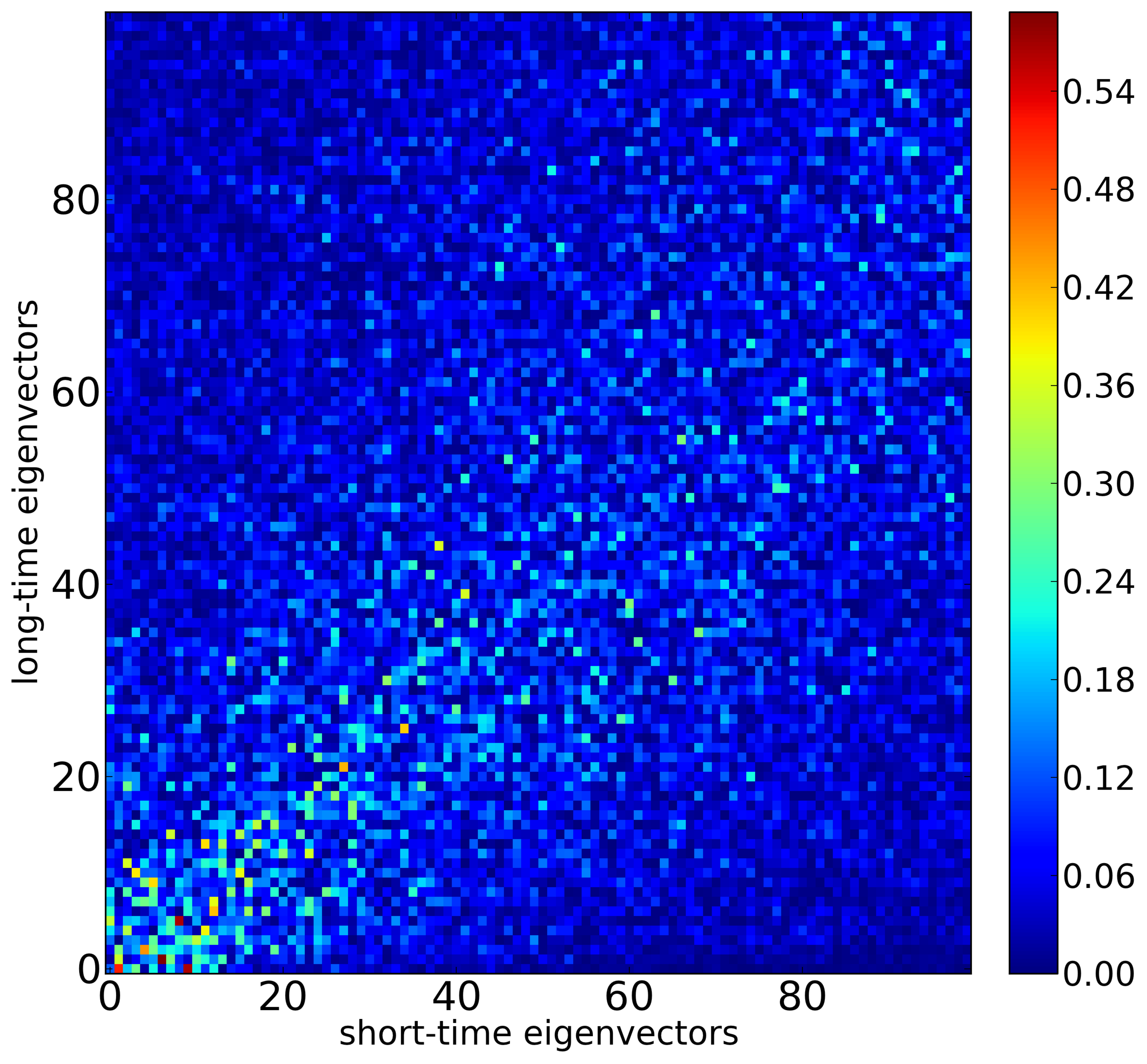}
  \caption{The overlap between the first 100 eigenvectors
    corresponding to the slowest fluctuational modes, ordered
    according to their eigenvalues, obtained from the first 1 ns
    (horizontal axis) and 10 ns (vertical axis) time frames. The
    accumulation along the lower diagonal indicates that the modal
    subspace spanned by slow eigenvectors retains its identity to a
    significant degree, with some amount of mixing between nearby
    modes.}
  \label{fig:mode_overlap}
\end{figure} 

Each time window considered was subjected to the same analysis. We
here present results for the first 1, 2, 5, and the full 10 ns of the
simulation. Fig.\ref{fig:mode_overlap} shows that the eigenvectors
corresponding to the 1 and 10 ns time frames, ordered with respect to
the amplitude of the corresponding eigenvalue, are in visible
agreement for the slow modes (lower left corner of the figure). This
observation is in line with earlier work which argues that the slow
modes retain their identity across different time scales, even if the
eigenvalue spectrum may
change~\cite{ma2005usefulness,pontiggia2007anharmonicity}. Early
identifyability of these most relevant modes points to the internal
consistency of our approach and supports our observation that, a mere
10 ns simulation is sufficient to extract meaningful information about
the biologically critical correlations in the dynamics which are
imprinted into the protein's complex structure.

\vspace*{1cm}
\section{Results}
\label{section:results}
\begin{figure}[h!]
  \centering
  \includegraphics[width=\textwidth]{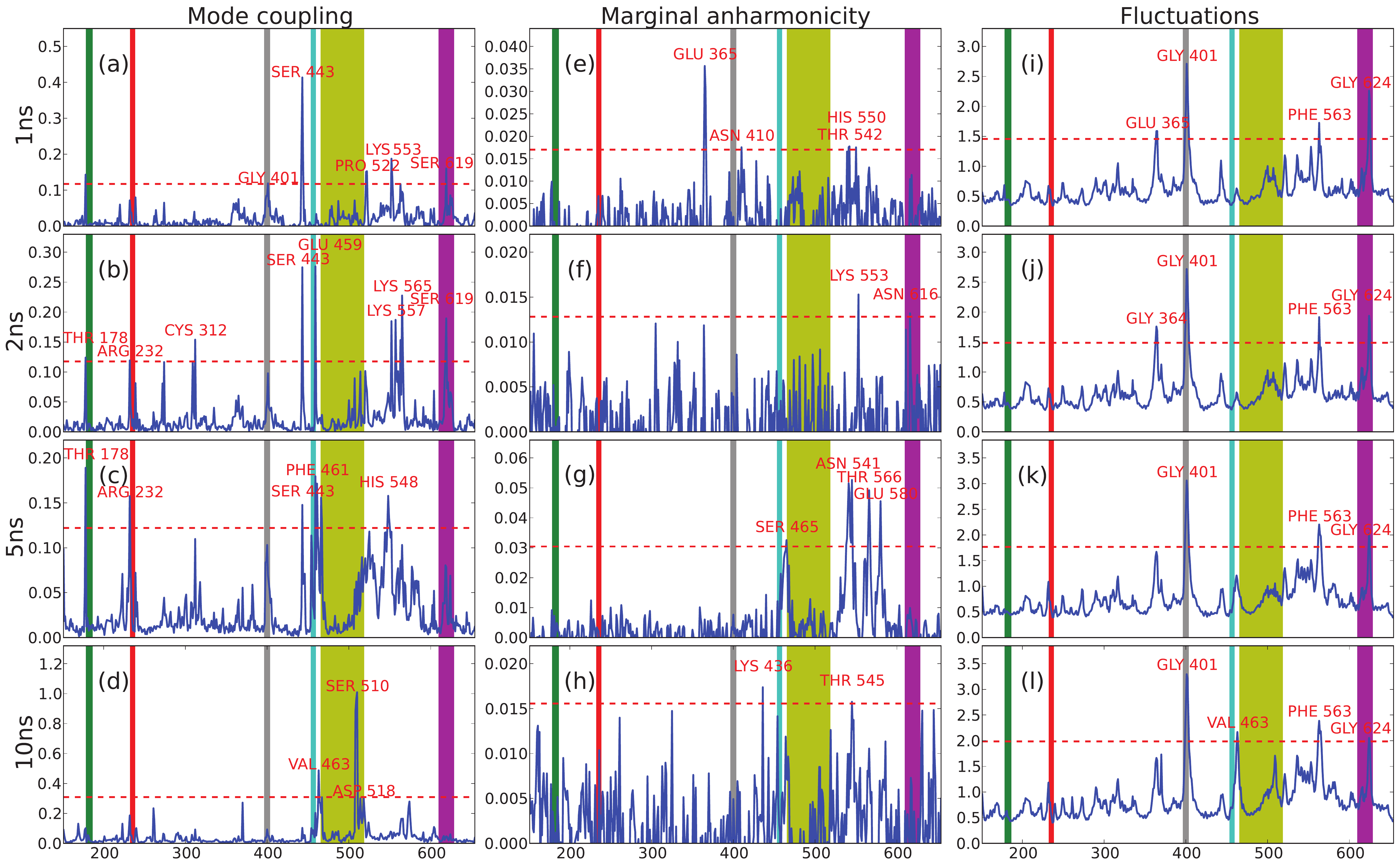}
  \caption{The residue scores obtained using
    Eqs.(\ref{eq:res_score_a},\ref{eq:res_score_b}) in the first 1 ns,
    2 ns, 5 ns, and 10 ns of the simulation, {\it w.r.t.}
    mode-coupling (first row), marginal anharmonicity (second row),
    and residue fluctuations (third row). Colored columns indicate
    functionally relevant regions, P-loop (green), Switch I (red),
    Myopathy loop (grey), Switch II (cyan), relay helix (yellow), and
    Loop-2 (purple), reported in the literature and described in
    Section~\ref{section:myosin_dynamics}. $3\sigma$ threshold is
    shown by the dashed line. The residue IDs of prominant peaks
    outside the $3\sigma$ margin are given in red.}
\label{fig:all_scores}
\end{figure} 

Fig.\ref{fig:all_scores} is a side-by-side comparison of residue
rankings obtained from the first 1 ns, 2 ns, 5 ns and 10 ns
simulations of the ligand-bound motor domain. The initial
configuration for both simulations was the structure PDB:1VOM and the
analysis was performed on the data collected after equilibration. In
each column, we consider three scoring schemes based on: (1)
mode-coupling, $S^{mc}_i$; (2) marginal anharmonicity, $S^{ma}_i$; (3)
mean residue fluctuation, $\sigma^2_i=\langle \Delta R_i^2 \rangle$.
In Fig.\ref{fig:all_scores}, the scores of the residues are shown for
each time frame and evaluation criterion, alongside the functional
regions indicated by different colored columns. We find that, the
high-scoring residues are marked by distinct peaks in mode-coupling
based ranking (more so than that based on fluctuation amplitudes),
while marginal anharmonicity is noisy and displays less selectivity
among residues. Upon comparing the magnitudes on the vertical axes in
the first two rows of Fig.\ref{fig:all_scores}, the relative weight of
mode coupling in an amino acid's fluctutational behavior is found to
be larger by an order of magnitude than that of marginal
anharmonicity. A similar observation was made on Crambin
earlier\cite{Kabakcioglu2010IOPscience}. Therefore, between the two
non-Gaussian contributions, mode-coupling appears as the dominant
factor in shaping the configurational landscape. Consistently, only
the residue-specific information gathered from mode-coupling
corrections yields a rank profile in significant agreement with
site-specific mutation data available for myosin II, as demonstrated
in Section~\ref{section:point_mutations}.

A closer inspection of mode-coupling based scores in different time
frames (first row in Fig.~\ref{fig:all_scores}) reveals an interesting
progression. We observe an increasing level of activity in the ligand
binding pocket during the first half of the simulation, but not
later. On the other hand, the contribution to the second half of the
simulation is mostly from the ends of the relay helix which is known
to mechanically couple the ligand- and actin-binding pockets. Such
shifts in activity in different time frames may be reminiscent of the
complex communication patterns established through energy transfer
between different vibrational modes in the
system~\cite{xie2000long,Piazza2009Longrange}. The scoring based on
residue fluctuation amplitudes alone (last row in
Fig.~\ref{fig:all_scores}) shows little difference between different
time frames, as one might expect.

Comparing first and last rows of Fig.\ref{fig:all_scores}, we observe
that there is some overlap between regions accentuated by mean
fluctuation amplitudes {\it vs.} by mode coupling. This is expected,
not only because our analysis derives from fluctuation data, but also
because some functional sites reside on flexible loop regions. We find
that the Myopathy loop and Loop-2, which is essential for actin
binding, yield a strong signal in both fluctuation and mode-coupling
based rankings. In contrast, loop regions in the ligand binding pocket
(P-loop, Switch I/II) do not fluctuate as much (presumably due to the
presence of the ligand), yet, they are still highlighted by mode
coupling.

\subsection{Comparison with point mutation data}
\label{section:point_mutations}
Score profiles in Fig.\ref{fig:all_scores}(a-d) further single out few
locations which are not in the immediate vicinity of the color coded
functional regions. For example, Ser-443, such a site selected by mode
coupling, coincides with the bent at the distal end of the long helix
between residues 411-440, known to promote the Myopathy loop to bind
actin~\cite{malnasi2005switch}. These may correspond to further
residues that are critical for protein function, for example relaying
information between the ligand- and actin-binding regions. For an
unbiased evaluation of all such instances, a higher resolution target
set is desirable. To this end, we performed a thorough literature
survey for residue-specific experimental data on myosin
II. Table~\ref{table:critical_residues} is a comprehensive list of
amino acids that we could gather for myosin II, which have been
experimentally verified (mostly through point mutations) to be
critical for its function. It is possible that, some of these are
important structural elements, say, required for folding, and not
necessarily critical in the sense of residing on a functional site or
maintaining allosteric communication. Nevertheless, this list
comprises a solid target set, free from theoretical considerations or
interpretations of structural data.

\vspace*{12pt}
\begin{table}[h!]
\begin{center}
\begin{tabular}{||p{5cm}|p{40pt}|||p{3cm}|p{100pt}||}
\hline
N483, F487, I499, F506, L508, I687, F692, F745 & Ref.~\cite{tsiavaliaris2002mutations} & D403, V405	& Ref.~\cite{Onishi2007Closer} \\ \hline
N464, C470,N472,Y473, N475, F481, E746 & Ref.~\cite{Ruppel1996Structurefunction} & Y494, W501 & Ref.~\cite{tsiavaliaris2002mutations, Patterson1996Coldsensitive}\\ \hline
D590, P591, L592, Q593 & Ref.~\cite{sasaki2000insertion} & S181 & Ref.~\cite{tang2007predicting} \\ \hline
E467, E586,G624, G740 & Ref.~\cite{Patterson1996Coldsensitive} & S236	 & Ref.~\cite{Frye2010Insights} \\ \hline
N233, S237, R238 & Ref.~\cite{shimada1997alanine} & E459 & Ref.~\cite{sasaki1998mutational,Ruppel1996Structurefunction}\\ \hline
I499, F692, R738 & Ref.~\cite{sasaki2003dictyostelium} & F482 & Ref.~\cite{tang2007predicting,Ito2003Requirement}\\ \hline
D454, G457, F458 & Ref.~\cite{sasaki1998mutational} & G680 & Ref.~\cite{tang2007predicting,Patterson1996Coldsensitive,Patterson1997Coldsensitive,Ito2003Requirement}\\ \hline
E531, P536, R562	& Ref.~\cite{Giese1997Phenotypically} & G691 & Ref.~\cite{Patterson1996Coldsensitive,Patterson1997Coldsensitive}\\ \hline

\hline

\end{tabular}
\caption{Amino acids that are experimentally verified to be critical
  for myosin II function. This list is used in the text as a target
  set for evaluating the relevance of various physical processes,
  namely, marginal anharmonicity, mode coupling, and fluctuation
  amplitudes, to protein's function.}
\label{table:critical_residues}
\end{center}
\end{table}

\begin{figure}[h!]
  \centering
  \includegraphics[width=\textwidth]{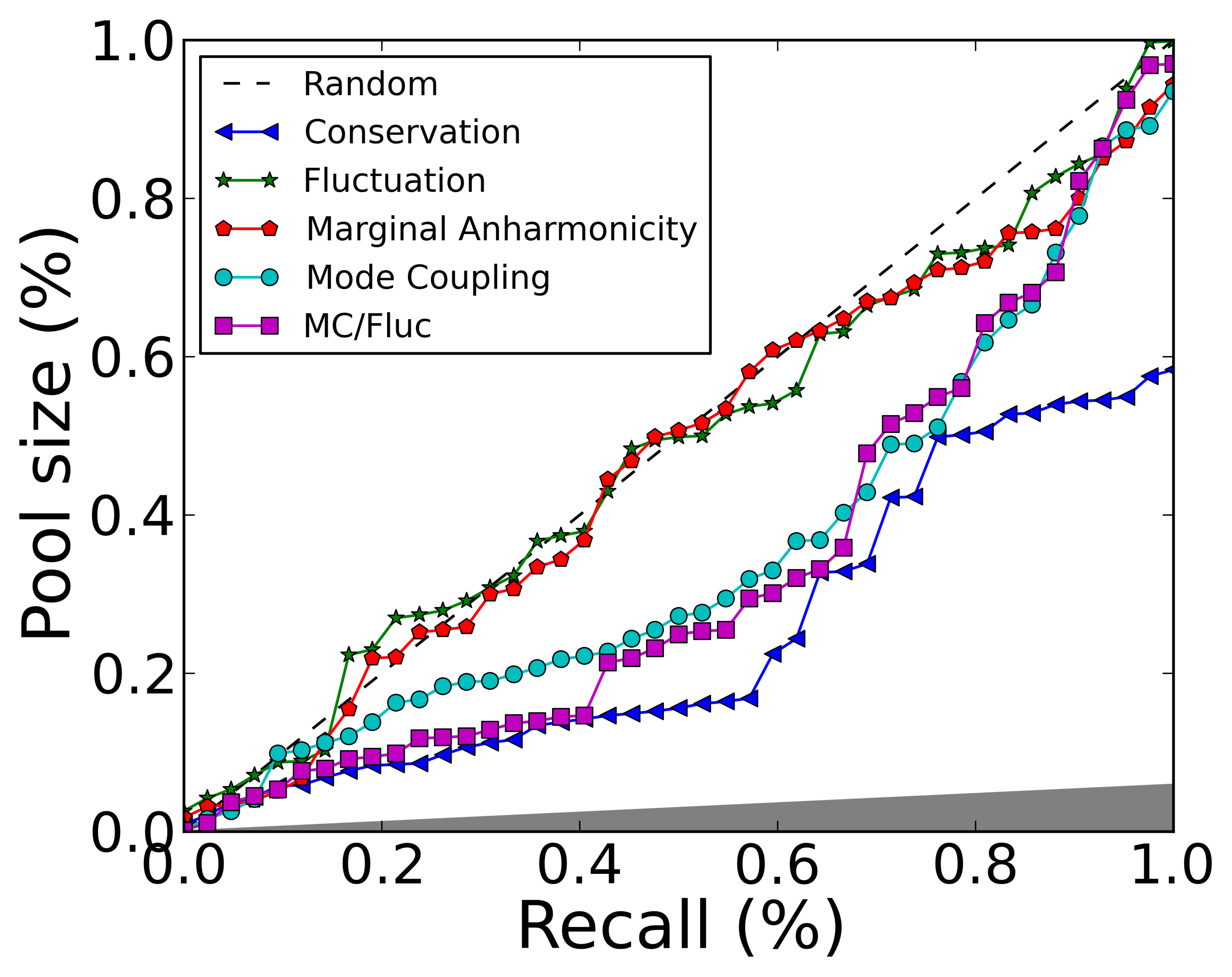}
  \caption{The performance of each ranking scheme considered in the
    paper is shown together with that of a random (dashed) and a
    perfect (upper edge of the forbidden black region) rank
    assignment. Each data point corresponds to the fraction of target
    residues captured out of a total of 43 (horizontal axis) by the
    top ranking residues covering a given percentage (vertical axis)
    of 730 residues in total.}
\label{fig:precision-recall}
\end{figure}

We next use a standard recall analysis to compare various ranking
methods (including random) against this target
set. Fig.\ref{fig:precision-recall} shows for each case, the
percentage of top residues that need to be considered (vertical axis)
in order to capture a certain fraction of the target set (horizontal
axis). For a random ordering of the residues, this curve is supposed
to lie on the diagonal, with fluctuations typically less than 10\% for
the current data. A perfect ordering which places the target residues
on top of the list would yield a curve following the upper edge of the
forbidden gray region at the bottom.

Upon inspecting Fig.~\ref{fig:precision-recall}, we observe that
fluctuations and marginal anharmonicity display no preference for the
target set. The conservation scores show the best correlation with the
experimentally determined targets. This is hardly surprising, since
experimental studies are in fact guided by conservation scores. The
key result of our study is the green curve representing the
mode-coupling based ordering. The clear deviation from the diagonal in
the downward direction demonstrates that coupling between vibrational
modes is an important physical mechanism for protein function and that
this information can be cast into a predictive tool by means of the
computational/analytical framework described above.

We also noticed that the scoring function $f_i \equiv
S^{mc}_i/\sigma^2_i$ which is also shown in
Fig.\ref{fig:precision-recall} (the curve labeled as ``MC/Fluc'') is
consistently better than $S^{mc}_i$ in terms of highlighting the
target set listed in Table~\ref{table:critical_residues}. We checked
that the general trend and the relative performances of different
scoring criteria are robust under different methods one might choose
while harnessing the information from different time windows (such as,
considering, for each amino acid, the maximum of a score among those
calculated from successive time intervals, instead of a single score
obtained from the full simulation). Note that, a gap builds in the
tail of Fig.\ref{fig:precision-recall} between conservation and
mode-coupling rankings. This discrepancy may be due to some residues
in the target list which are relevant for the folding process
(therefore also have high conservation scores), but not the functional
dynamics where we expect mode coupling to play a role. Further
investigations on myosin II and other proteins are in progress to
confirm this hypothesis, as well as to verify the generality of the
present approach.

Figs.\ref{fig:precision-recall} presents a direct validation of the
fact that, coupling between vibrational modes, as formulated here, is
a significant physical mechanism underlying myosin II's functional
dynamics. This is the main message we wish to convey in this
paper. Since the existing data on myosin II is far from being
exhaustive, it is likely that further experiments will identify more
essential residues for this protein. The method introduced here may
also help in these future endevours as a new guide for target
selection (for example, Ser-443 appears to be a promising
candidate). The distribution of top amino acids with high
mode-coupling scores shown in Fig.~\ref{fig:FunctionalSites} is
further encouraging in this respect. One observes that, the top
scoring residues are not randomly distributed across the
structure. Rather, they accumulate around the core of the motor
domain, in visible agreement with the distribution of the residues in
the target set in Table~\ref{table:critical_residues}.

\begin{figure}[h]
  \includegraphics[width=\textwidth]{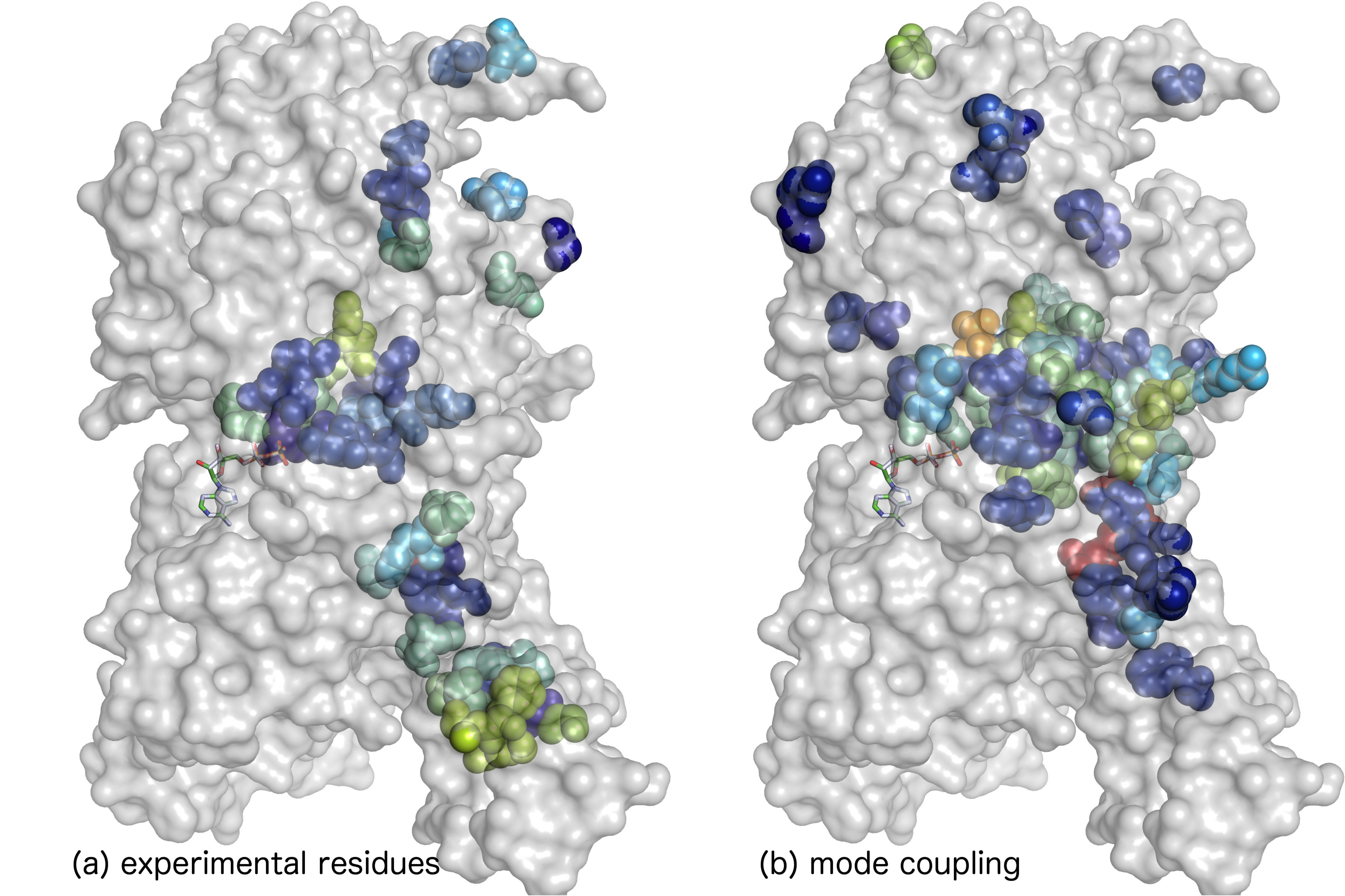}
  \caption{A side-by-side comparison of the locations of residues in
    (a) the experimentally determined target set and (b) top 10\% w.r.t. 
    mode-coupling scores. Hot colors indicate higher scores.}
  \label{fig:FunctionalSites}
\end{figure}

\section{Discussion}
\label{section:discussion}
It is generally accepted that the coupling between different
vibrational modes is an important physical mechanism driving
correlated functional dynamics in proteins, particularly in the
context of allosteric communication. However, this wisdom generated
little input for experimentalists so far. We here propose an
analytical/computational framework where the ``noninteracting'' limit
is composed of already anharmonic modes (consistent with the observed
slow modes in proteins). Mode coupling is then defined as everything
that falls outside the best possible description of the
configurational distribution (obtained from simulations) as a function
factorizable in such modes. The information content of the
mode-coupling contribution obtained by this operational definition can
be utilized to highlight certain locations on the protein. Despite the
fact that the MD simulations are much shorter than the time required
to observe functionally relevant dynamics, we show here that these
locations correlate with critical residues/regions obtained from
experiments on the motor domain of myosin II.

Our work simultaneously confirms the relevance of mode coupling to
function and proposes a new computational tool for predicting
functionally critical locations on proteins. However, considering the
multitude of factors that contribute to the evolutionary design these
complex machines, it is difficult to imagine our approach (or any
other non-hybrid, {\it ab initio} method) to singlehandedly yield
sufficiently accurate predictions. Assuming our tests currently in
progress on several other proteins yield favorable results, a more apt
use of the present computational framework would be to employ it as a
module in a multifaceted prediction algorithm that seeks consensus
between complementary approaches. Several such tools are publicly
available~\cite{prymula2011catalytic}.

\section*{Acknowledgments}
We thank The Scientific and Technological Research Council of Turkey
(T\" UB\. ITAK) for the grant MFAG-113F092 awarded to support the
application of the present framework on a number of allosteric
proteins. Part of the computing resources used during this work were
provided by the National Center for High Performance Computing of
Turkey (UYBHM) under grant number 4001752012.

\section*{References}
\bibliographystyle{unsrt}
\bibliography{manuscript_pb}{}
\end{document}